\documentclass[sigconf]{acmart}
\mathchardef\mhyphen="2D
\usepackage{multirow}
\usepackage{subfigure}
\AtBeginDocument{%
  \providecommand\BibTeX{{%
    \normalfont B\kern-0.5em{\scshape i\kern-0.25em b}\kern-0.8em\TeX}}}

\setcopyright{acmcopyright}
\copyrightyear{2021}
\acmYear{2021}
\acmDOI{10.1145/3459637.3482170}

\acmConference[CIKM '21] {Proceedings of the 30th ACM International Conference on Information and Knowledge Management}{November 1--5, 2021}{Virtual Event, Australia.}
\acmBooktitle{Proceedings of the 30th ACM Int'l Conf. on Information and Knowledge Management (CIKM '21), November 1--5, 2021, Virtual Event, Australia}
\acmPrice{15.00}
\acmISBN{978-1-4503-8446-9/21/11}

\begin{document}
\fancyhead{}
\title{Relation-aware Heterogeneous Graph for User Profiling}

\author{Qilong Yan\textsuperscript{\rm 1,2}\footnotemark[1], Yufeng Zhang\textsuperscript{\rm 1}\footnotemark[1], Qiang Liu\textsuperscript{\rm 1,2}, Shu Wu\textsuperscript{\rm 1,2}\footnotemark[2] and Liang Wang\textsuperscript{\rm 1,2} \\}

\affiliation{%
\institution{\textsuperscript{\rm 1}Center for Research on Intelligent Perception and Computing, Institute of Automation, Chinese Academy of Sciences \\
\textsuperscript{\rm 2}School of Artificial Intelligence, University of Chinese Academy of Sciences \\}
\country{\{qilong.yan,yufeng.zhang\}@cripac.ia.ac.cn \\
\{qiang.liu,shu.wu,wangliang\}@nlpr.ia.ac.cn \\}
}

\renewcommand{\shortauthors}{Qilong Yan, Yufeng Zhang, Qiang Liu, Shu Wu and Liang Wang}
\renewcommand{\authors}{Qilong Yan, Yufeng Zhang, Qiang Liu, Shu Wu and Liang Wang}

\begin{abstract}
  User profiling has long been an important problem that investigates user interests in many real applications. Some recent works regard users and their interacted objects as entities of a graph and turn the problem into a node classification task. However, they neglect the difference of distinct interaction types, e.g. user clicks an item v.s. user purchases an item, and thus cannot incorporate such information well. To solve these issues, we propose to leverage the relation-aware heterogeneous graph method for user profiling, which also allows capturing significant meta relations. We adopt the query, key, and value mechanism in a transformer fashion for heterogeneous message passing so that entities can effectively interact with each other. Via such interactions on different relation types, our model can generate representations with rich information for the user profile prediction. We conduct experiments on two real-world e-commerce datasets and observe a significant performance boost of our approach.
\end{abstract}




\maketitle

\renewcommand{\thefootnote}{\fnsymbol{footnote}}
\footnotetext[1]{Equal contribution.} 
\renewcommand{\thefootnote}{\arabic{footnote}}

\renewcommand{\thefootnote}{\fnsymbol{footnote}}
\footnotetext[2]{Corresponding author.} 
\renewcommand{\thefootnote}{\arabic{footnote}}

\section{Introduction}
Nowadays, users from all kinds of applications have been producing an ocean of data as they browse the internet. Such data potentially contain valuable information, such as user's interest, trait, and behaviour pattern, for providing them with personalised services. The scenario is typically named user profiling. Taking \citep{rao2010classifying} as an example, they first propose to predict user's gender, age, area, and political tendency by modelling their twitters from social networks. Recent works expand the task purpose to a broader scope, including occupation \citep{zhao2019appusage2vec}, geolocation \citep{rahimi2018semi}, ideology \citep{xiao2020timme}, and race \citep{preoctiuc2018user}. 

\begin{figure}[t]
  \includegraphics[width=0.45\textwidth]{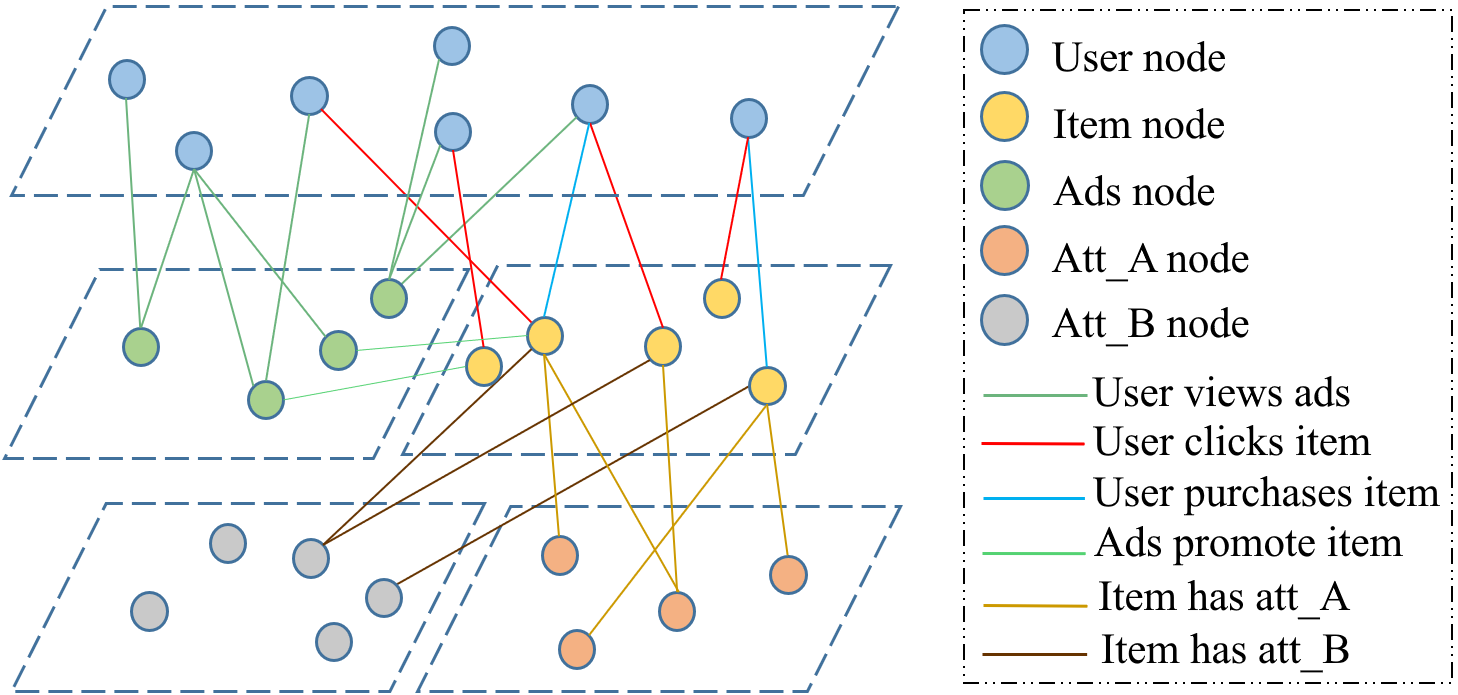}
  \caption{The heterogenerous graph with multiple types of entities and relations for user profiling. Best view in colour.}
  \label{fig:graph}
  \vspace{-5mm}
\end{figure}

In user profiling, an intuitive way is to model the user's interaction behaviour with graphs. Despite the success of traditional deep learning approaches \citep{rao2010classifying,nguyen2013old,farnadi2018user}, graph methods are highlighting their advantages on non-euclidean relations in such tasks \citep{Wu:2019ke,Yu:2020dn,zhang2020every,zhang2020personalized,li2020dynamic,Zhang:2021wv,zhang2021graph}. \citet{rahimi2018semi,chen2019semi,xiao2020timme} regard users with co-relation (like co-purchase in e-commerce) as a graph with entities and hierarchically pump the heterogeneous information up from the attribute with graph attention networks. In addition to the interaction, the semantic of entities is also important. For example, items usually possess the description of their category and brand, and advertisements have that of their sponsor and campaign, which are unified as side information \citep{liu2021non}. \citet{chen2019semi} apply the words of the title as entities to represent the side information.

\begin{figure*}
  \includegraphics[width=0.8\textwidth]{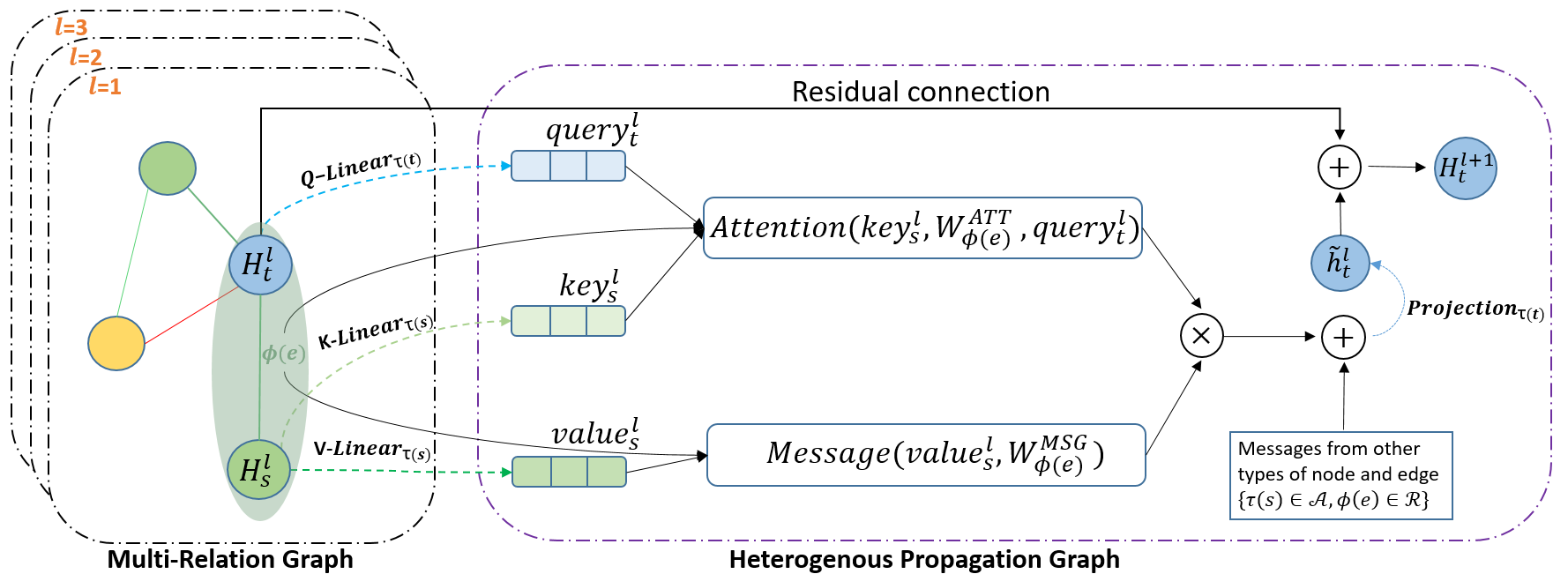}
  \caption{The overall architecture of the relation-aware heterogeneous graph network. Suppose an edge $e$ links a source node $s$ and a target node $t$, denote its meta relation as $\langle\tau(s), \phi(e), \tau(t)\rangle$. The representation $H_t$ of the user is obtained by collecting the neighbourhood messages.}
  \label{fig:model}
  \vspace{-3mm}
\end{figure*}

However, two problems remain untouched in existing works. First, previous studies oversimplify the relations between entities and merely count on the binary association (with and without interaction). In real-world scenarios, users generally interact with other objects with multiple behaviours, e.g. they can click, like, or purchase items on a typical shopping platform. We argue that different behaviours may contribute to different intentions and the degree of favour. A `click' is naturally a less strong association than a `like' or `purchase'. Second, the side information is insufficiently considered so far. Only titles of the item are deemed as attributes, whereas other types, such as categories and brands, are dismissed. Simply integrating more types of side information into the graph may not be sound since they expect different semantic space when doing attention operations. Actual experiments have proved that they do not produce satisfactory results.

The core issue mentioned above is that most works are based on a single type of entity and relation, but more information needs to be uncovered by different types. To that end, we propose Relation-aware Heterogeneous Graph Network for user profiling (RHGN) that can model multiple relations on the heterogeneous graph. In contrast to the single relation graph by previous approaches, we adopt a graph with various relations between different types of entities, as illustrated in Figure \ref{fig:graph}. We also design a heterogeneous graph propagation network that aggregates information from multiple sources. A transformer-like multi-relation attention \citep{vaswani2017attention, hu2020heterogeneous} is employed to learn the importance between nodes and reveal the meta-relation significance on the graph. We validate our model on two real-world user profiling datasets. The experimental result shows that our approach significantly advances the prediction for user profiles. The main contributions of this paper are as follows:
\begin{itemize}
\item We first propose the heterogeneous graph with multiple types of relations and entities for user profiling.

\item We adopt a heterogeneous graph propagation network to acquire heterogeneous information from multiple sources.
    
\item We use heterogeneous multi-relation attention to automatically reveal the meta-relation significance on the graph.
\end{itemize}

\section{Relation-aware Heterogeneous Graph Networks}

In this section, we formulate the user profiling problem and introduce our approach to address the multi-source information extraction on the heterogeneous graph. As illustrated in Figure \ref{fig:model}, the model consists of three segments: \textit{Neighbourhood Message Passing}, \textit{Multi-Relation Attention}, and \textit{Target State Update}.

\subsection{Problem Statement}

Given a collection of users' behaviours and properties, user profiling aims to predict their labels, such as age and gender. We present the entities and relations as a directed heterogeneous graph $\mathcal{G}=(\mathcal{V}, \mathcal{E}, \mathcal{A}, \mathcal{R})$, where nodes $v\in \mathcal{V}$ and edges $e\in \mathcal{E}$ can be mapped into their types by function $\tau(v): V\rightarrow \mathcal{A}$ and $\phi(e): E\rightarrow \mathcal{R}$, respectively. In the case of e-commerce, $\tau(v)$ typically has types `user', `item', `advertisement', and `attribute', and $\phi(e)$ has types `click', `purchase', and `has\_attribute', as illustrated in Figure \ref{fig:graph}. Suppose an edge $e$ links a source node $s$ and a target node $t$, denote its meta relation as $\langle\tau(s), \phi(e), \tau(t)\rangle$. The meta relation generally reflects different interaction intentions between entities. 

Note that in real situations, labels for the user are usually limited. Semi-supervised learning is thus required by using a large number of unlabelled data. 

\subsection{Neighbourhood Message Passing}

To obtain the representation $H$ of users, it is necessary to collect their neighbourhood messages, that is, the items they have interacted with. They, to a large extent, exhibit one's interest and concern. Similarly, items should learn from their neighbourhood users and attributes. Formally, for a triplet of a target node $t$, its neighbour node $s \in N(t)$, and an edge relation $e=(s, t)$, we calculate the message to pass in the $l$-th layer as:
\begin{equation}
    M(s, e, t) =  \mathop{\parallel}\limits_{i \in [1, h]} M\mhyphen head^i(s, e, t) \\
    \vspace{-2mm}
\end{equation}
\begin{equation*}
    M\mhyphen head^i(s, e, t) = F^i_M \left(H^l[s] \right)W^{MSG}_{\phi(e)} \\
\end{equation*}
where $\parallel$ denotes the concatenation operation, $F^i_M$ is the $i$-th multi-head linear function, $h$ is the number of heads, and $W^{MSG}_{\phi(e)}$ is a matrix that projects the message into a relation-dependent space. By keeping a distinct $W^{MSG}_{\phi(e)}$ for each meta relation $\langle\tau(s), \phi(e), \tau(t)\rangle$, our model can better distinguish different intentions between the source and the target with various types of interaction, e.g. click and purchase. 

\begin{table*}[t]
    \centering
    \caption{The comparison between various models on the two datasets.}
    \vspace{-3mm}
    \begin{tabular}{ccccc|cccc}
    \toprule
    \multirow{2}{*}{Model} & \multicolumn{4}{c}{JD-dataset} & \multicolumn{4}{c}{Alibaba-dataset} \\
    \cmidrule(l){2-9}
    & \multicolumn{1}{c}{Gender-Acc} & \multicolumn{1}{c}{Gender-F1} & \multicolumn{1}{c}{Age-Acc} & \multicolumn{1}{c}{Age-F1} & \multicolumn{1}{c}{Gender-Acc} & \multicolumn{1}{c}{Gender-F1} & \multicolumn{1}{c}{Age-Acc} & \multicolumn{1}{c}{Age-F1} \\ \midrule
    GCN & 40.60 & 28.88 & 51.83 & 17.80 & 78.81 & 45.48 & 32.14 & 14.34 \\
    GAT & 77.52 & 75.85 & 51.45 & 25.52 & 79.09 & 47.75 & 34.89 & 16.70 \\
    RGCN & 60.30 & 56.40 & 45.10 & 17.10 & 78.00 & 69.96 & 36.94 & 22.28 \\
    HGCN & 80.00 & 79.33 & 52.82 & 25.77 & 80.97 & 63.82 & 41.55 & 26.78 \\
    (+info) & 78.10 & 77.52 & 52.30 & 23.74 & 80.56 & 63.27 & 41.23 & 26.12 \\ 
    HGAT & 80.20 & \textbf{79.40} & 51.70 & 19.46 & 79.89 & 64.27 & 36.72 & 21.76 \\
    (+info) & 78.74 & 78.05 & 51.40 & 17.13 & 79.12 & 63.70 & 35.16 & 20.58 \\ \midrule
    RHGN & \textbf{80.44} & 79.18 & \textbf{54.70} & \textbf{33.95} & \textbf{83.00} & \textbf{77.73} & \textbf{43.80} & \textbf{29.60} \\
    \bottomrule
    \end{tabular}
    \label{tab:result}
    \vspace{-3mm}
\end{table*}

\subsection{Multi-Relation Attention}

Like many sorts of research demonstrated \citep{ rahimi2018semi,chen2019semi}, not all neighbourhood messages are necessarily essential for the target node. For example, the model may need to pay more attention to which item the user bought rather than which advertisement they viewed. We thus adopt an attention weight to rescale the significance of each message. In formal, we project the source node and the target node into a Key vector and a Query vector, respectively, and measure their similarity as:
\begin{equation}
    \vspace{-3mm}
    \alpha(s, e, t) = \mathop{\rm softmax}\limits_{\forall s\in N(t)}\left(\mathop{\parallel}\limits_{i \in [1, h]}\alpha\mhyphen head^i\left(s, e, t\right)\right) \\
\end{equation}
\begin{equation*}
    \alpha\mhyphen head^i(s, e, t) = \left(K^i(s)W^{ATT}_{\phi(e)}Q^i(t)^T\right)\cdot\frac{1}{\sqrt{d}}\\
    \vspace{-1mm}
\end{equation*}
\begin{equation*}
    K^i(s) = F^i_K(H^l[s])
    \vspace{-1mm}
\end{equation*}
\begin{equation*}
    Q^i(t) = F^i_Q(H^l[t])
\end{equation*}
where $F^i_K$ and $F^i_Q$ are the $i$-th multi-head linear function for the Key vector and the Query vector, respectively, $W^{ATT}_{\phi(e)}$ is a projection matrix, and $d$ is the dimension of the vector. The architecture resembles Transformer \citep{ vaswani2017attention}, and $\sqrt{d}$ is also used to smooth the dot product of the Key vector and the Query vector. However, the vanilla Transformer calculates the dot product with the same set of parameters for all inputs, which does not consider the effect of multiple associations. The additional weight $W^{ATT}_{\phi(e)}$ here can help the model reassign attentions according to different meta relations.

\subsection{Target State Update}

After propagating the messages and their attentions to the target node, we assemble them to update the target node embedding. We define the update formulation as:
\begin{equation}
    H^{l+1}[t] = F_{map}\left(\sigma(\hat{H}^l[t])\right) + H^l[t]
    \vspace{-2mm}
\end{equation}
\begin{equation*}
    \hat{H}^l[t] = \mathop{\oplus}\limits_{\forall s\in N(t)}\left(\alpha\left(s, e, t\right)\cdot M\left(s, e, t\right)\right)
\end{equation*}
where $F_{map}$ is a linear function that maps the message back to the target feature distribution, and $\sigma$ denotes the activation function. As the attention $\alpha(s, e, t)$ is normalised by the softmax procedure ($\sum_{\forall s\in N(t)}\alpha(s, e, t) = 1_{h\times1}$), it can be directly applied to the message $M(s, e, t)$ without affecting its distribution. Through stacking the graph layer and the residual connection, each node can reach $L$-hop neighbours. For example, a user can receive messages from other users who may share similar interest and behaviour, though they are not connected. 

\subsection{Training}

The final step involves classifying the user representation in the last layer into profile labels. Formally, we employ a single linear classifier and optimize the model with cross-entropy as:
\begin{equation}
    \mathcal{L} = -\sum\limits_{t\in \mathcal{V}'}\sum\limits^{P}_{p=1} Y_{tp}{\rm log}(Z_{tp})
    \vspace{-2mm}
\end{equation}
\begin{equation*}
    Z = {\rm softmax}\left(F_{out}(H^L[t])\right)
\end{equation*}
where $\mathcal{V}'$ denotes the labelled user node set, $P$ denotes the total number of profile labels, and $Y$ denotes the ground truth.

\section{EXPERIMENTS}

In this section, we conduct experiments on two real-world datasets to evaluate our proposed method.

\subsection{Datasets}

To examine the actual performance of our proposed method, we select two public large-scale user profiling datasets in real scenes: JD-dataset\footnote{https://github.com/guyulongcs/IJCAI2019\_HGAT} and Alibaba-dataset\footnote{https://tianchi.aliyun.com/dataset/dataDetail?dataId=56}, two of the most popular e-commerce portals in China. For each dataset, the heterogeneous graphs are extracted with multiple relations between users, items (or advertisements), and attributes. In consistency with \citep{chen2019semi}, we use the user’s gender and age as the label of their profiles.  In the JD-dataset, users and items have `click' and `purchase' relations, and attributes include four category descriptions of items. In the Alibaba-dataset, users and items have four relations - `click', `purchase', `favorite', and `shopping cart'; users and advertisements have `view' relations; items and advertisements have 
`promotion' relations; and attributes have three types of basic information of advertisements, including category ID, campaign ID, and sponsor ID.

\subsection{Baselines}

We consider both classical and state-of-the-art graph methods as baselines:
(1) \textbf{GCN} \citep{kipf2016semi} and \textbf{GAT} \citep{velivckovic2017graph} are two representative and strong baselines on many tasks, that are based on homogeneous graphs and do not take mutiple types of relations and entities into account;
(2) \textbf{RGCN} \citep{schlichtkrull2018modeling} refers to Relational GCN. It splits the graph into several sub-graphs according to different types of relations and uses parallel GCN layers for each sub-graph; 
(3) \textbf{HGCN} and \textbf{HGAT} \citep{chen2019semi} are two state-of-the-art methods. They regard the node as heterogeneity. However, they lack the distinction between relation types. They also dismiss some important side information;
(4) We further extend HGCN and HGAT with \textbf{(+info)} to investigate whether they gain benefit from more side information. We add side information under the same attention distribution of their model.

\begin{table}[t]
    \centering
    \caption{Ablation study on the JD-dataset}
    \vspace{-3mm}
    \small
    \begin{tabular}{ccccc}
        \toprule
         Model & Gender-Acc & Gender-F1 & Age-Acc & Age-F1  \\ \midrule
         RHGN & 80.44 & 79.18 & 54.70 & 33.95 \\
         w/o U-I relations & 78.99 & 77.56 & 54.62 & 31.01 \\
         w/o I-A relations & 75.95 & 74.24 & 52.69 & 30.23 \\
        \bottomrule
    \end{tabular}
    \label{tab:ablation}
    \vspace{-5mm}
\end{table}

\subsection{Experimental Setup }

We implement our RHGN in the PyTorch framework for efficient GPU computation. In the experiment, we randomly split labelled users into a training set, a validation set, and a test set with the ratio 75 : 12.5 : 12.5 \citep{qiu2018deepinf,chen2019semi}. The embedding for users and items is randomly initialized, whereas that for attributes is initialized by their content via Fasttext \citep{armand2017bag}. We adopt the grid-search strategy to find the optimal parameter combination for our model. The entity-level aggregation network has two layers with the hidden dimension in \{32, 64, 128, 256\}. The number of heads in multi-head attention is searched in \{1, 2, 4, 8, 16\}. All models are optimized via the AdamW optimizer with the One Cycle Learning Rate Scheduler. The learning rate, weight decay, and mini-batch size are set to 0.001, 0.01, 512, respectively. We use GELU \citep{hendrycks2016gaussian} as our activation function. The implementation of all baselines follow their original paper.


There are two node classification tasks: the gender prediction (binary classification task) and the age prediction (multi-class classification task). We evaluate the models with Accuracy and Macro-F1 \citep{wu2019neural,chen2019semi}, which are widely used in user profiling problems.

\subsection{Results}

Table \ref{tab:result} displays the experimental results of different methods on the two datasets. We observe that our model significantly boost the performance of most tasks. In particular, our model presents an averagely higher performance gain on the Alibaba dataset than that on the JD dataset. It is reasonable because the Alibaba dataset contains more diverse interaction behaviours, which carries richer user intentions. By modelling distinct meta relations, our model can intrinsically extract more information than the baselines. 

The result also shows that HGCN and HGAT outperform vanilla GCN and GAT, implying that the task benefits from the heterogeneous node types. Nevertheless, they do not improve further by incorporating more side information. It is probably because they project different types of side information into the same distribution so that they cannot discriminate the impact of each. 

Overall, the experiment indicates that it is necessary to consider multiple types of meta relations in user profiling, and our approach can leverage such information to provide better services. 

\begin{figure}[t]
    \centering
    \subfigure[]{
    \includegraphics[width=0.45\textwidth]{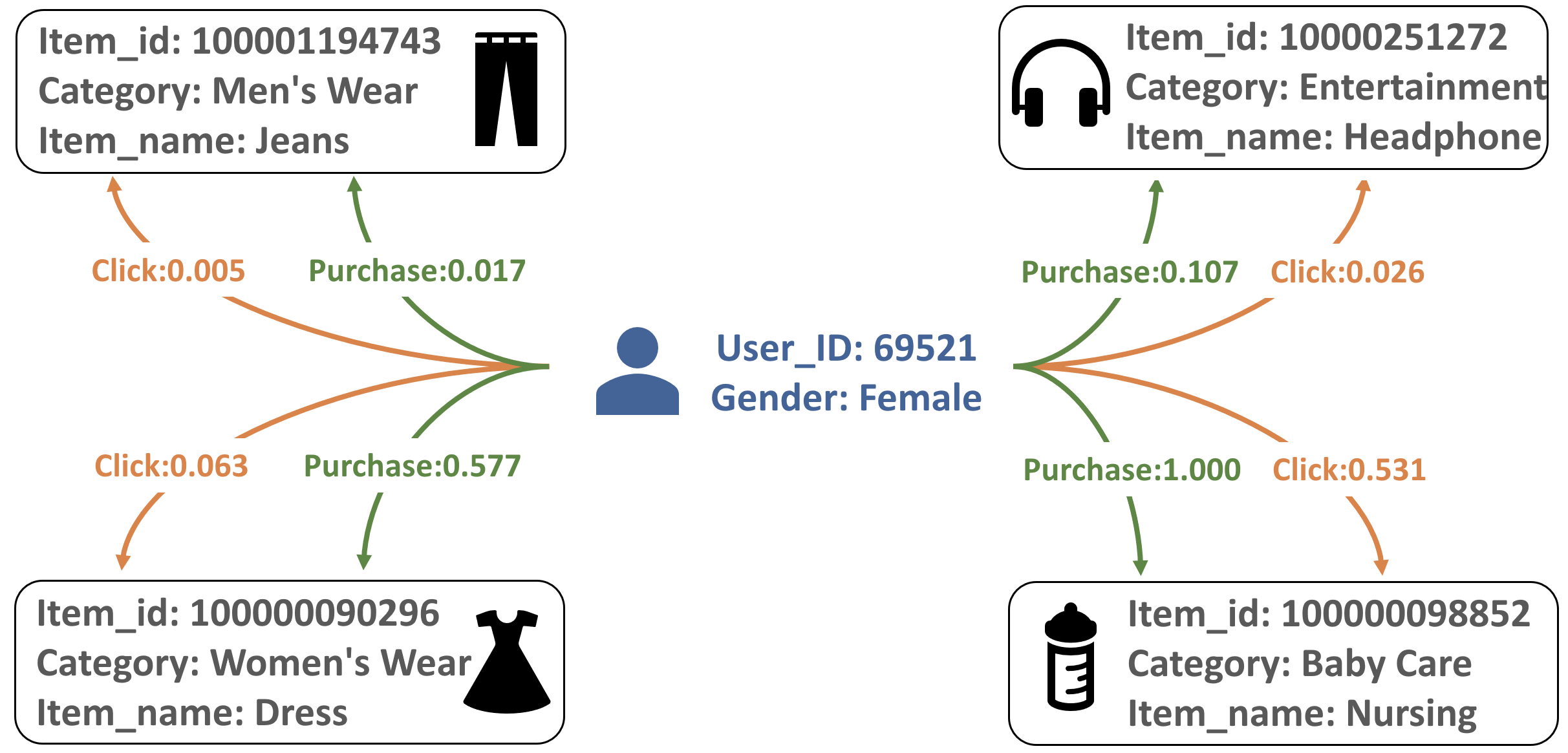}}
    \vspace{-3mm}
    \subfigure[]{\includegraphics[width=0.45\textwidth]{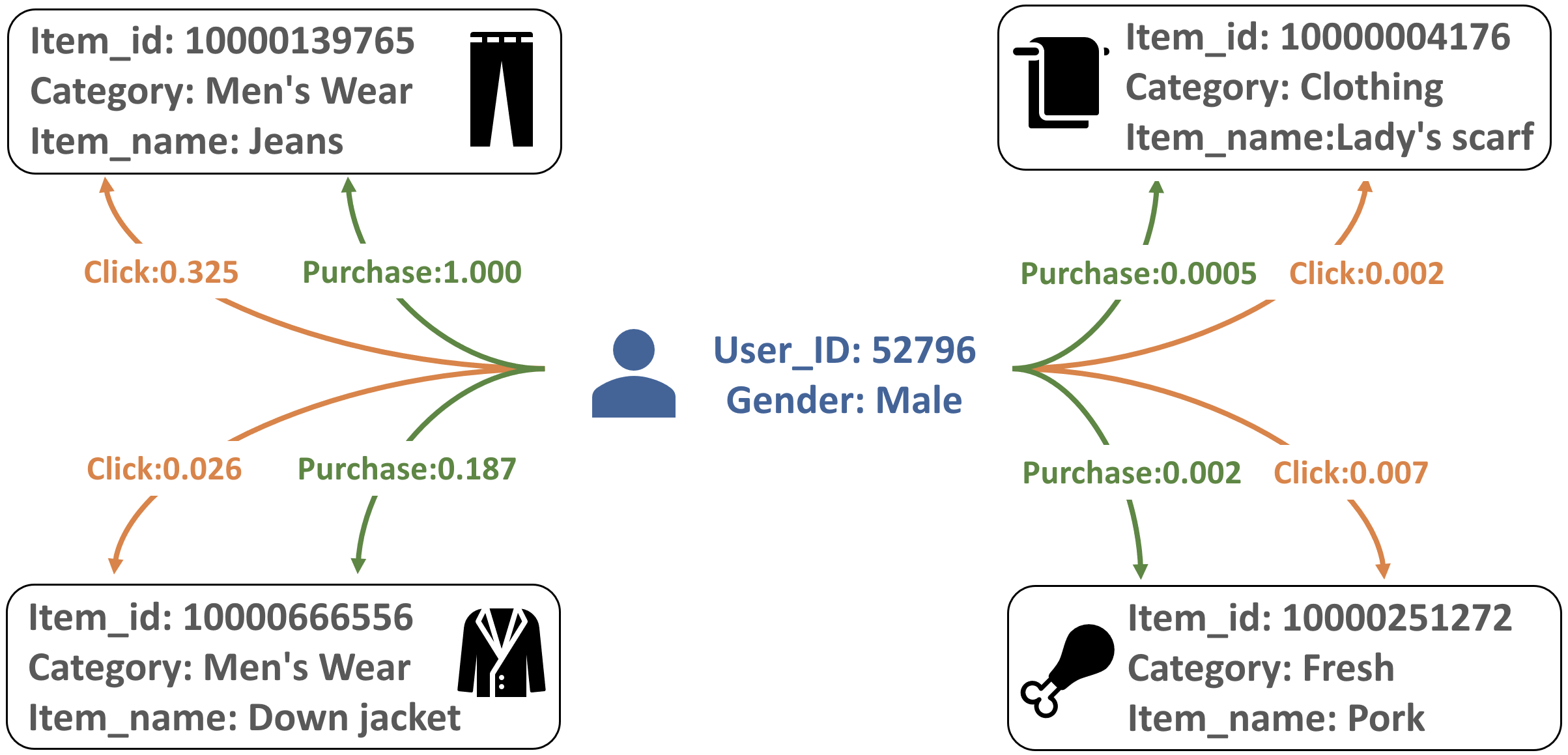}}
    \vspace{-2mm}
    \caption{An example from the JD-dataset. Visualizing the significance of distinct relation types with different items for opposite genders.
    }
    \label{fig:visualize}
    \vspace{-5mm}
\end{figure}

\subsection{Ablation Study}

To investigate the individual effectiveness of user-item multi-relation and item-attribute (side information) multi-relation, we carry out ablation study experiments on them. Specifically, we modify RHGN by consolidating user-item attention (w/o U-I relations) and item-attribute relations (w/o I-A relations), respectively. As demonstrated in Table \ref{tab:ablation}, the result shows both individuals can improve the performance compared with original model, suggesting that either multi-relation can contribute to the task. In addition, the item-attribute relation yields a higher performance influence than the user-item relation. It is a plausible phenomenon since side information contains category semantics that can reflect user intentions.

\subsection{Case Study}

To understand how the meta relation impacts the prediction for user profiles, we visualize the attention score between two users (a male and a female) and their interacted items, as illustrated in Figure \ref{fig:visualize}. According to different user genders, the attention score exhibits different significances in terms of the item category and interactive relation. It is worth noting that the `click' relation of some gender-oriented items are more biased than the `purchase' relation of some neutral items.

\section{Conclusion}

In this paper, we proposed a heterogeneous graph with multiple entities and relations for user profiling. We also adopted a relation-aware heterogeneous graph network to learn the meta relation significance on such a graph. Through experiments on real large datasets, we found incorporating more types of entities and relations is generally beneficial for capturing user's intentions and predicting their profile labels. Further studies demonstrate the interpretability of
RHGM with different meta relation attention weights.

\section*{Acknowledgments}

This work is supported by National Key Research and Development Program (2019QY1601, 2019QY1600), National Natural Science Foundation of China (61772528).


\bibliographystyle{ACM-Reference-Format}
\bibliography{reference}


\end{document}